\documentclass[sigconf]{acmart}
\AtBeginDocument{%
  }

\copyrightyear{2026}
\acmYear{2026}
\setcopyright{cc}
\setcctype{by}
\acmConference[MM '26]{Proceedings of the 34th ACM International Conference on Multimedia}{November 10--14, 2026}{Rio de Janeiro, Brazil}
\acmBooktitle{Proceedings of the 34th ACM International Conference on Multimedia (MM '26), November 10--14, 2026, Rio de Janeiro, Brazil}
\acmDOI{10.1145/3767308.3834908}
\acmISBN{979-8-4007-2213-4/2026/11}

\usepackage{multirow}
\usepackage{pifont}
\usepackage{colortbl}
\usepackage{subcaption}

\newlength\savewidth\newcommand\shline{\noalign{\global\savewidth\arrayrulewidth
\global\arrayrulewidth 1pt}\hline\noalign{\global\arrayrulewidth\savewidth}}

\begin{document}

\title{Hear to See: Discerning Stateful Listening for Audio-Visual Instance Segmentation}

\author{Leiye Liu}
\email{leiyeliu@mail.dlut.edu.cn}
\orcid{0009-0003-9400-2428}
\affiliation{%
  \institution{Dalian University of Technology}
  \city{Dalian}
  \state{Liaoning}
  \country{China}
}

\author{Miao Zhang}
\email{miaozhang@dlut.edu.cn}
\correspondingauthor
\orcid{0000-0002-7972-7047}
\affiliation{%
  \institution{Dalian University of Technology}
  \city{Dalian}
  \state{Liaoning}
  \country{China}
}

\author{Jiahong Jiang}
\email{jiahongjiang@mail.dlut.edu.cn}
\orcid{0009-0006-2447-1968}
\affiliation{%
  \institution{Dalian University of Technology}
  \city{Dalian}
  \state{Liaoning}
  \country{China}
}

\author{Jingjing Li}
\email{jingjingli.cmu@gmail.com}
\orcid{0000-0003-0811-4988}
\affiliation{%
  \institution{Carnegie Mellon University}
  \city{Pittsburgh}
  \state{Pennsylvania}
  \country{USA}
}

\author{Jialong Zhong}
\email{jialongzhong@mail.dlut.edu.cn}
\orcid{0009-0001-7813-8486}
\affiliation{%
  \institution{Dalian University of Technology}
  \city{Dalian}
  \state{Liaoning}
  \country{China}
}

\author{Kai Peng}
\email{happypk@mail.dlut.edu.cn}
\orcid{0009-0005-9280-1693}
\affiliation{%
  \institution{Dalian University of Technology}
  \city{Dalian}
  \state{Liaoning}
  \country{China}
}

\author{Tingwei Liu}
\email{tingweiliu@mail.dlut.edu.cn}
\orcid{0009-0005-3558-6528}
\affiliation{%
  \institution{Dalian University of Technology}
  \city{Dalian}
  \state{Liaoning}
  \country{China}
}

\author{Wei Ji}
\email{wei.ji@yale.edu}
\orcid{0000-0003-4059-5902}
\affiliation{%
  \institution{Yale University}
  \city{New Haven}
  \state{Connecticut}
  \country{USA}
}

\author{Yongri Piao}
\email{yrpiao@dlut.edu.cn}
\correspondingauthor
\affiliation{%
  \institution{Dalian University of Technology}
  \city{Dalian}
  \state{Liaoning}
  \country{China}
}

\author{Huchuan Lu}
\email{lhchuan@dlut.edu.cn}
\affiliation{%
  \institution{Dalian University of Technology}
  \city{Dalian}
  \state{Liaoning}
  \country{China}
}

\renewcommand{\shortauthors}{Leiye Liu et al.}

\begin{abstract}
  Audio-visual instance segmentation (AVIS) requires accurately identifying and tracking individual sounding objects with pixel-level masks. Existing methods struggle to match overlapping acoustic events with visual instances and handle asynchronous audio-visual dynamics. Therefore, two critical questions arise: how can a model establish precise correspondence between overlapping sound sources and visual instances, and how can a model maintain robust tracking when audio and visual signals are temporally misaligned?This paper proposes Hear to See (H2S), addressing these challenges through two mechanisms. The Acoustic-Semantic Projector (ASP) disentangles mixed audio and establishes hierarchical correspondence from semantic to spatial domains. The Asynchronous Dynamics Modulator (ADM) adaptively adjusts state transitions via audio-modulated Mamba, prioritizing current information during dynamic variations and maintaining continuity in stable periods.Experiments on AVISeg show H2S achieves SOTA performance, attaining 48.54 mAP with a COCO pretrained ResNet50 and surpassing the previous by 7.8\%. The code will be open-sourced once the paper is accepted. The source code will be publicly available at \url{https://github.com/leiyeliu/H2S}.
\end{abstract}




\maketitle

\section{Introduction}
\label{sec:intro}

Humans possess an exceptional ability to integrate sensory information across modalities such as vision and audition. This capacity forms the foundation of complex cognition. Among various sensory modalities, the combination of vision and sound is particularly crucial, as videos naturally couple auditory and visual cues. Enabling machines to jointly interpret what is seen and what is heard is fundamental for advanced video understanding \cite{avvp, pavel}, human computer interaction \cite{caver}, and autonomous perception \cite{agdm}.

Driven by this goal, research on audio–visual learning has progressed from audio–visual sound localization (AVSL) \cite{tovavel, vavl} to audio–visual segmentation (AVS) \cite{COMBO,AVSegFormer,AVSBG,AVSBench}, which focuses on all sounding objects. However, these methods still struggle to separate different sound sources of the same category and to model long-term audio–visual relations in complex environments. To overcome these limitations, audio–visual instance segmentation (AVIS) \cite{avis} has been introduced. It distinguishes each sounding object with pixel-level masks, enabling fine-grained multimodel understanding.

\begin{figure}
    \centering
    \includegraphics[width=.95\columnwidth]{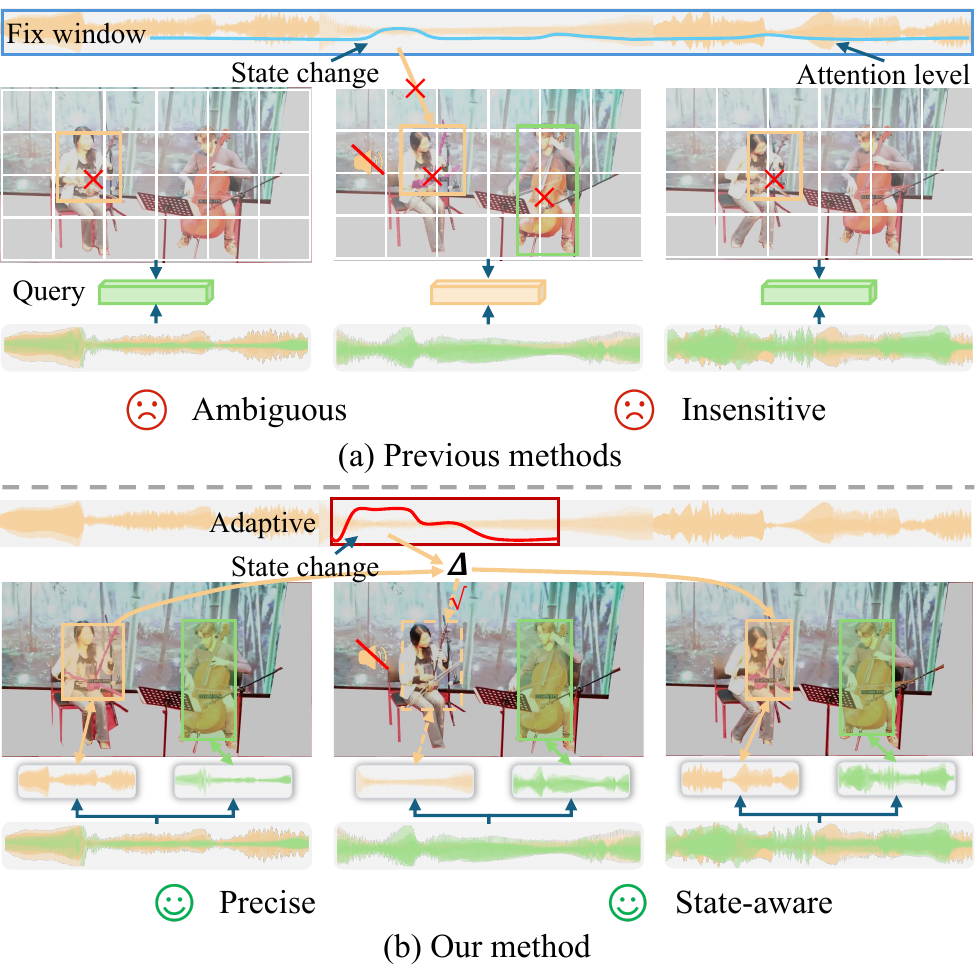}
    \vspace{-14pt}
    \caption{Comparison between previous methods and ours. (a) Previous methods suffer from ambiguous correspondence with mixed audio, and their fixed-window modeling causes insensitive tracking of asynchronous sounding states. (b) Our method achieves precise matching by disentangling acoustic sources, and enables adaptive state-aware tracking by dynamically modulating state transitions via $\Delta$.}
    \vspace{-8pt}
    \Description{The figure illustrates the comparison between previous methods and ours.}
    \label{fig:motivation}
\end{figure}

The key of AVIS is to establish a pathway from "hearing" to "seeing", i.e., acoustic cues must effectively guide visual reasoning to segment sounding object instances. Achieving this goal requires addressing two problems, as shown in Figure \ref{fig:motivation}.
The first stems from the heterogeneity between auditory and visual modalities. Visual information represents object instances across spatial dimensions, while acoustic information superimposes multiple concurrent sound events into a single temporal channel. This heterogeneity requires models to possess the capability to parse independent acoustic events from the mixed single channel and establish accurate correspondence between these events and spatially visual objects. To address this, current approaches \cite{AVSegFormer, avis,VCT} employ an intermediate representation, such as queries or prototypes, to bridge acoustic and visual modalities. However, this design inevitably loses fine-grained acoustic details that distinguish individual sound events during the conversion of acoustic signals to abstract representations for cross-model matching. Without these details, it becomes difficult to establish correspondence between modalities.

The second concerns modeling the asynchronous dynamics between audio and video streams. In real-world scenarios, audio and video streams exhibit dynamic and time-varying coupling relationships. Visual instances alternate between sounding and silent states, leading to completely asynchronous information density and states across the two modalities along the time. This requires the model to possess dynamic state awareness capabilities: perceiving their sounding versus silent states in real time and maintaining tracking after sound cessation. However, existing methods \cite{avis} employ grid-based temporal processing mechanisms, such as fixed-window attention or consistent memory aggregation, to associate instances across frames. These approaches process all audio-visual instances within equal-length temporal windows, regardless of whether they are sounding or silent. This limitation causes frequent tracking failures when sound and visual presence are asynchronous.

This paper proposes a novel framework, Hear to See audio-visual instance segmentation model (H2S), which consists of two core components: the Acoustic-Semantic Projector (ASP) and the Asynchronous Dynamics Modulator (ADM). Specifically, ASP disentangles mixed audio into independent acoustic streams, preserving fine-grained acoustic details crucial for instance discrimination. Subsequently, we employ a hierarchical correspondence mechanism to extract cluster centers from acoustic streams and visual features, establishing associations at the semantic level. These associations are then mapped back to the spatial domain, enabling precise correspondence between audio and vision. By doing so, the model ensures both clear identification of individual instances and accurate cross-model matching.
Furthermore, we propose ADM based on an audio-dynamically modulated Mamba. We feed visual instances into Mamba, where the $\Delta$ parameters capture visual dynamics. Building on this, we introduce audio dynamics into Mamba to modulate the visual $\Delta$ parameters. This design enables ADM to adjust its update strategy based on audio-visual dynamics: when dynamics exhibit large variations, the model prioritizes current information to capture abrupt changes; when dynamics remain stable, the model emphasizes historical continuity to preserve long-term associations. Ultimately, this mechanism ensures instance identity continuity and tracking robustness even in complex scenarios.

Benefiting from acoustic disentanglement and dynamic state modeling, H2S achieves state-of-the-art performance on the AVISeg dataset. With a COCO pretrained ResNet50 backbone, our method attains a mAP of 48.54, surpassing the previous best approach by 7.8\%. These results demonstrate the effectiveness of our framework in achieving accurate cross model correspondence and robust temporal modeling for complex audio-visual scenarios, alongside zero-shot generalization and stable tracking of asynchronous events.

\section{Related Work}
\label{sec:RW}

\begin{figure*}
    \centering
    \includegraphics[width=.95\linewidth]{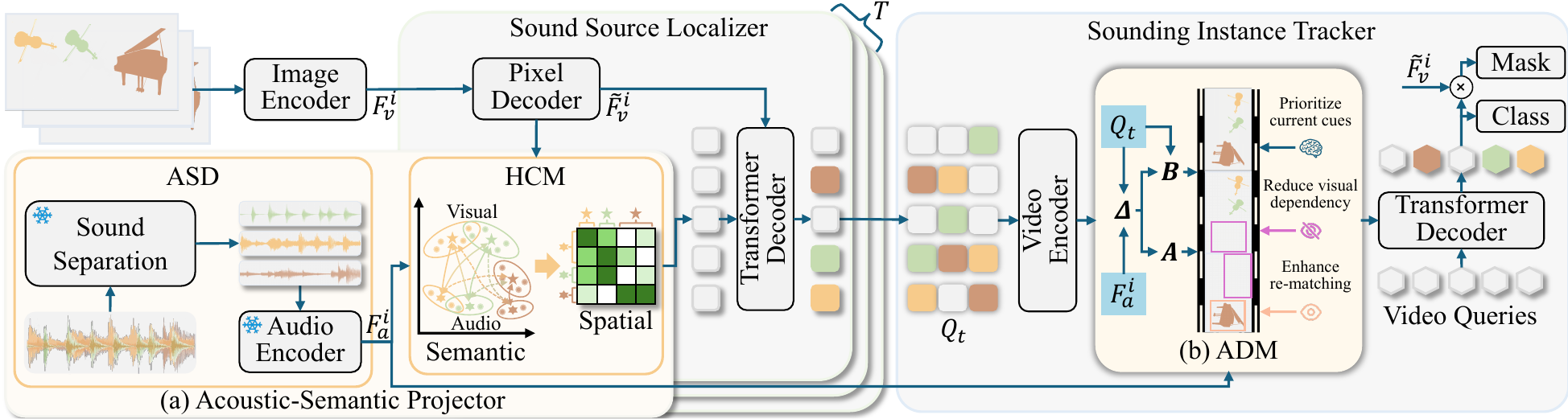}
    \vspace{-10pt}
    \caption{Overall architecture of the proposed H2S. (a) The proposed Acoustic-Semantic Projector (ASP), consisting of Audio Source Disentanglement (ASD) and Hierarchical Correspondence Mechanism (HCM), establishes precise cross-model correspondence between audio and visual features in complex acoustic environments. (b) The proposed Asynchronous Dynamics Modulator (ADM) enables robust instance tracking and consistent identification under asynchronous audio-visual dynamics.}
    \vspace{-10pt}
    \Description{The figure illustrates the overall architecture of the proposed H2S model.}
    \label{fig:method}
\end{figure*}

\subsection{Audio-Visual Segmentation}
Audio-visual segmentation (AVS) \cite{AQFormer,AVSegFormer,pk,pk2} aims to identify and segment visual regions corresponding to sounding objects, requiring both accurate localization and precise spatial delineation. Early works leverage audio as guidance for visual attention. AQFormer \cite{AQFormer} treats audio features as object queries to capture sound-related regions and facilitate cross-model interaction. AVSegFormer~\cite{AVSegFormer} further improves this interaction with bidirectional conditional fusion using audio-visual mixers and decoders. COMBO~\cite{COMBO} introduces a Bilateral Fusion Module and an Adaptive Inter-Frame Consistency Loss to enhance temporal coherence. SelM~\cite{SelM} uses a selective mechanism to filter noisy features and constrain spatiotemporal representations corresponding to sounding instances.  
Other approaches explore contrastive and prototype-guided strategies. CAVP~\cite{CAVP} employs supervised audio-visual contrastive learning to enhance embedding quality, while VCT~\cite{VCT} generates prototype-guided queries to capture spatial regions and audio event categories.  

Building upon these efforts, AVISM~\cite{avis} first introduces audio-visual instance segmentation, extending AVS from pixel-level to instance-level prediction. It proposes the AVIS dataset and develops a baseline model. The baseline localizes sound sources in each frame and tracks sounding objects using window attention to propagate audio-visual temporal information. This enables more detailed and dynamic understanding of complex audio-visual scenes.

\subsection{Visual Mamba}
State space models have rapidly emerged as a compelling alternative to Transformers due to their linear complexity and effective modeling of long-range dependencies. Many researchers have introduced them into vision tasks and achieved promising progress \cite{vim,vmamba,defmamba}. In the multimodel domain, existing methods can be broadly categorized into two primary paradigms. The first focuses on designing sophisticated scanning inherent strategies to reconcile structural differences between modalities \cite{AVS-Mamba,zhong2026adasurvmamba}. These methods aim to achieve better cross-model features alignment.
The second paradigm revolves around the internal structure of Mamba, such as cross-model gating and content fusion. These methods \cite{LEAF-Mamba,MSFMamba} primarily design interactions by exchanging SSM parameters like $\mathbf{B}$ and $\mathbf{C}$ or activation branches. The goal is to enable one modality to dynamically gate or enrich the feature representation of another modality at each time step. While this approach is effective for creating joint representations, its internal mechanism remains homogeneous across different contexts. In other words, cross-model interaction is primarily focuse on the content fusion level.

In contrast, our proposed ADM introduces a fundamentally new perspective. We argue that for the AVIS task, the key lies in how the modeling process itself should dynamically adapt to the varying states of multiple modalities. This approach represents a novel exploration of Mamba in the multimodel domain.

\section{Method}
\label{sec:method}

\subsection{Preliminaries}

State Space Models (SSMs), such as S4~\cite{s4} and Mamba~\cite{mamba}, are structured sequence models that integrate the core principles of recurrent and convolutional architectures. They achieve linear or near-linear computational complexity with respect to sequence length, making them highly efficient for long-range sequence modeling. These models originate from continuous-time dynamical systems and define a mathematical mapping from an input sequence \(u(t) \in \mathbb{R}^L\) to an output sequence \(y(t) \in \mathbb{R}^L\) through a hidden state \(h(t) \in \mathbb{R}^N\), where \(t\) denotes time.

Formally, an SSM can be expressed as a continuous-time Ordinary Differential Equation (ODE)~\cite{mamba}:
\vspace{-3pt}
\begin{equation}
\begin{aligned}
    h'(t) &= \mathbf{A}h(t) + \mathbf{B}u(t), \\
    y(t) &= \mathbf{C}h(t),
\end{aligned}
\end{equation}
\vspace{-6pt}

\noindent where \(h(t)\) is the hidden state, \(h'(t)\) is its temporal derivative, and \(u(t)\) and \(y(t)\) denote the input and output, respectively. The matrices \(\mathbf{A} \in \mathbb{R}^{N \times N}\), \(\mathbf{B} \in \mathbb{R}^{N \times 1}\), and \(\mathbf{C} \in \mathbb{R}^{1 \times N}\) represent the state transition, input projection, and output projection.

To effectively apply SSMs in discrete-time learning frameworks, the continuous system must first be discretized into a sequence-to-sequence formulation. Both S4~\cite{s4} and Mamba~\cite{mamba} employ a Zero-Order Hold (ZOH)~\cite{mamba} discretization scheme, introducing a timescale parameter \textbf{\(\Delta\) that controls the state transition speed and determines how quickly information propagates through time, Intuitively, a larger $\Delta$ forces the model to focus on the current input to capture abrupt state transitions, while a smaller $\Delta$ preserves historical context during stable periods.}. The continuous parameters \(\mathbf{A}\) and \(\mathbf{B}\) are subsequently discretized into \(\bar{\mathbf{A}}\) and \(\bar{\mathbf{B}}\) as follows:
\vspace{-3pt}
\begin{equation}
\begin{aligned}
    \bar{\mathbf{A}} &= \exp(\Delta \mathbf{A}), \\
    \bar{\mathbf{B}} &= (\Delta \mathbf{A})^{-1} \big(\exp(\Delta \mathbf{A}) - \mathbf{I}\big) \, \Delta \mathbf{B}, \\
    h_t &= \bar{\mathbf{A}} h_{t-1} + \bar{\mathbf{B}} u_t, \\
    y_t &= \mathbf{C} h_t.
\end{aligned}
\label{eq:2}
\end{equation}
\vspace{-6pt}

Although both S4 and Mamba follow the same principle, Mamba introduces input-dependent parameterization through the Selective Scan (S6) mechanism. Specifically, the parameters \(\Delta \in \mathbb{R}^{b \times L \times D}\), \(\mathbf{B} \in \mathbb{R}^{b \times L \times N}\), and \(\mathbf{C} \in \mathbb{R}^{b \times L \times N}\) are conditioned on the input sequence \(u_{t} \in \mathbb{R}^{b \times L \times D}\), where \(b\) denotes the batch size, \(L\) is the sequence length, and \(D\) represents the feature dimension.

\subsection{Overall Architectures}
Given a video sequence, we partition it into non-overlapping video clips $\{ X_{t} \}_{t=1}^{T}$ and their corresponding audio streams $\{ A_{t} \}_{t=1}^{T}$, where $T$ denotes the sequence length. As illustrated in Figure \ref{fig:method}, following the prior workes \cite{mask2former, vita}, our method takes $\{ X_{t} \}_{t=1}^{T}$ and $\{ A_{t} \}_{t=1}^{T}$ as inputs and aims to segment instance masks of sounding objects $\{M_{t,i}\}_{t=1, i=1}^{T, N_t}$, where $N_t$ represents the number of instances at frame $t$. 
The video signal is first encoded by a pretrained backbone network, yielding multi-scale video features $\{ F_{v}^{i} \}_{i=1}^{4}$. Here, $F_{v}^{i} \in \mathbb{R}^{B \times T \times D \times H_{i} \times W_{i}}$, where $B$ denotes the batch size, $D$ is the feature dimension, and $H_{i}, W_{i}$ represent the spatial dimensions at stage $i$. To preserve fine-grained details crucial for instance discrimination, the audio signal is fed into the ASD (as described in Section \ref{331}) for sound source separation and feature extraction, producing audio features $\{ F_{a}^{i}\}_{i=1}^{N_a}$. Here, $F_{a}^{i} \in \mathbb{R}^{B \times T \times D}$, where $N_a$ represents the number of separated sound sources.
These two features are then passed into $T$ frame-level sound source localizers to localize sounding objects. The video features are input to a pixel decoder for multi-scale interaction, yielding refined features $\{\tilde{F}_{v}^{i}\}_{i=1}^{4}$. Subsequently, these refined features and audio features are jointly fed into the HCM (as described in Section \ref{332}) for cross-model correspondence and fusion. Frame-level object queries are then extracted using a transformer decoder.
Finally, $\{\tilde{F}_{v}^{i}\}_{i=1}^{4}$, queries from all frames, and $\left\{ F_{a}^{i}\right\}_{i=1}^{N_a}$ are sent to the video-level sounding instance tracker for matching and tracking sounding instances. After passing through the video encoder, the queries and $\left\{ F_{a}^{i}\right\}_{i=1}^{N_a}$ are input to the ADM (as described in Section \ref{34}), which incorporates dynamic variations from both video and audio modalities to enhance the model's reasoning capability. A transformer decoder then extracts video-level queries to predict classes, which are combined with the $\{\tilde{F}_{v}^{i}\}_{i=1}^{4}$ to predict masks.

\subsection{Acoustic-Semantic Projector}
To enable the model to identify individual instances clearly while ensuring accurate cross-model matching, we propose the ASP. This model comprises two components: the Audio Source Disentanglement (ASD) and the Hierarchical Correspondence Mechanism (HCM). The ASD focuses on separating audio sources, while the HCM establishes multi-model relationships between audio and video. Note that the ASD is executed once, whereas the HCM operates multiple times within the frame-level sound source localization.

\subsubsection{\bf Audio Source Disentanglement} \hfill \\
\label{331}
\indent Audio perception confronts a fundamental challenge. When multiple sounding instances vocalize simultaneously, their sounds convolve with each other or with unknown acoustic reverberation functions due to the physical nature of sound. They mix in acoustic sensors in an inseparable manner, forming a single entangled audio signal. Such audio makes it difficult to distinguish different instances when establishing cross-modal correspondences with video, and this one-to-many relationship is inherently problematic. Some methods attempt to split a single audio feature into multiple query representations using cross-attention, but this approach lacks explicit sound source supervision and relies solely on the network's self-learning. In contrast, our method employs MixIT \cite{mixit}, a sound source separation model pretrained on YFCC100M \cite{mixit}, to separate audio sources. This approach explicitly disentangles individual sound sources. The separated sources are then converted into mel-spectrograms and fed into a pretrained VGGish \cite{vggish} model for feature extraction, yielding audio features $\{ F_{a}^{i}\}_{i=1}^{N_a}$. Note that both the source separation model and feature extraction model remain frozen during this process, which is executed only once.

\begin{figure}
    \centering
    \includegraphics[width=.9\linewidth]{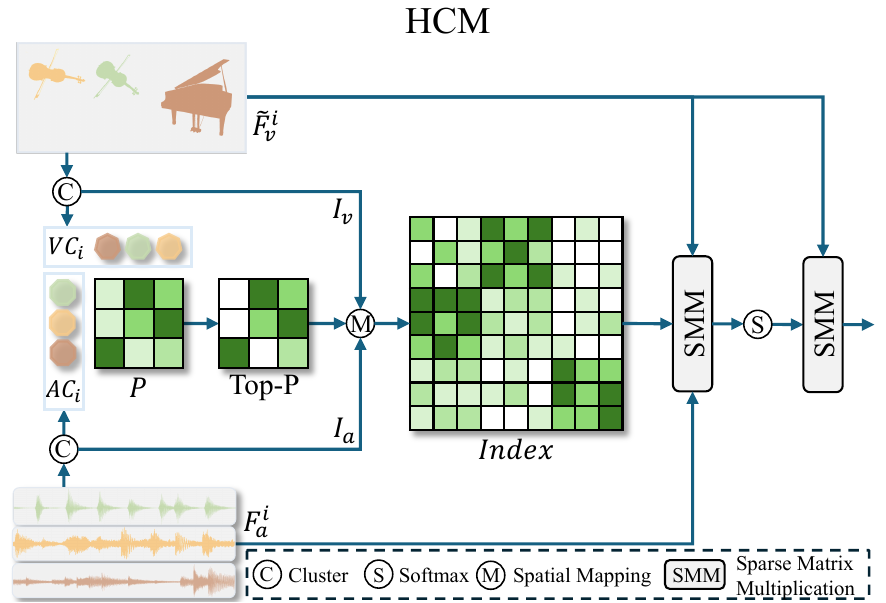}
    \vspace{-8pt}
    \caption{Structure of the Hierarchical Correspondence Mechanism (HCM). For clarity, the figure shows only three clusters each for audio and video, and three audio feature streams. In practice, the number of clusters and streams exceeds three.}
    \vspace{-14pt}
    \label{fig:model1}
    \Description{The figure illustrates the structure of the Hierarchical Correspondence Mechanism (HCM).}
\end{figure}

\begin{figure*}
    \centering
    \includegraphics[width=.9\linewidth]{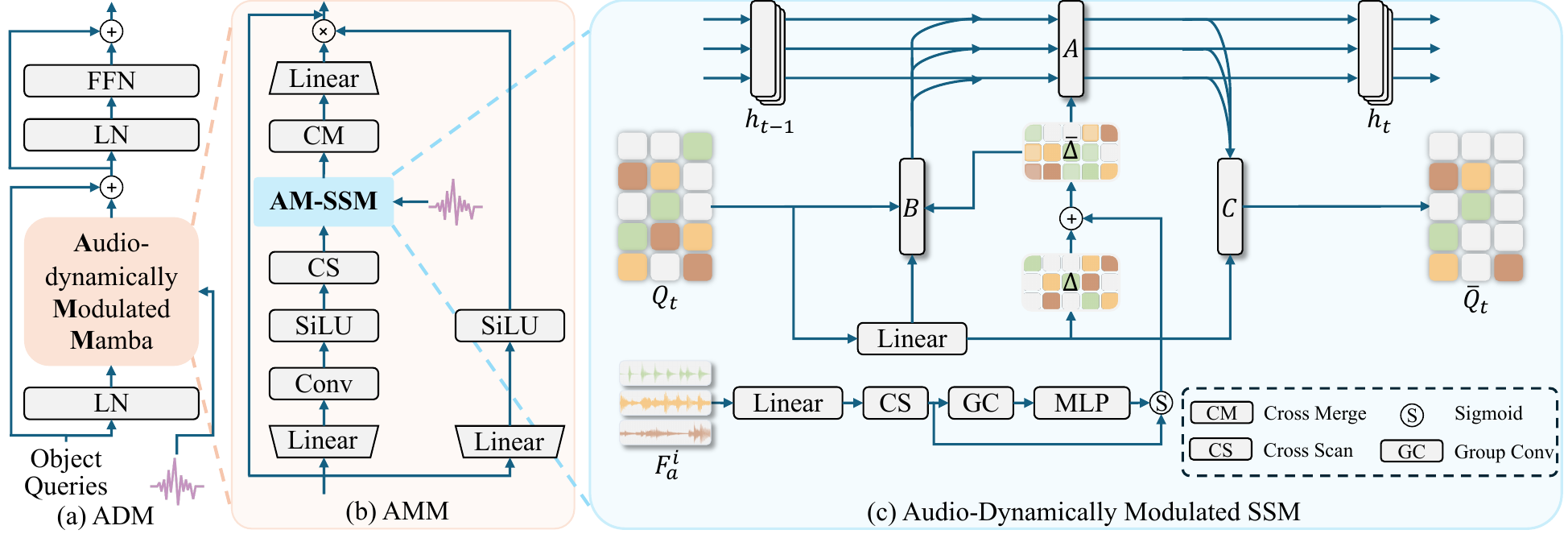}
    \vspace{-10pt}
    \caption{Structure of the Asynchronous Dynamics Modulator (ADM). (a) Overall structure of ADM. (b) Structure of Audio-Dynamically Modulated Mamba (AMM). (c) Data flow in Audio-Dynamically Modulated SSM (AM-SSM).}
    \vspace{-10pt}
    \label{fig:model2}
    \Description{The figure illustrates the structure of the Asynchronous Dynamics Modulator (ADM).}
\end{figure*}

\subsubsection{\bf Hierarchical Correspondence Mechanism} \hfill \\
\label{332}
\indent The heterogeneity between audio and video modalities results in distinct feature characteristics: video features prioritize spatial representation, while audio features emphasize temporal summarization. However, both modalities share a common semantic ground, as their semantic information jointly refers to the scene constructed in the video. To address this, we propose the HCM, which first establishes connections at the semantic semantic level, then maps these connections back to the spatial domain to obtain precise spatial correspondences, and performs multi-model interaction. The detailed structure is illustrated in the Figure \ref{fig:model1}.

Given the refined video features $\{\tilde{F}_{v}^{i}\}_{i=1}^{4}$ and audio features $\{ F_{a}^{i}\}_{i=1}^{N_a}$, we first extract features from the top three stages. We then apply k-means clustering to both video and audio features separately, mapping them into a shared semantic space to obtain audio cluster $\{AC_{i}\}_{i=1}^{C_{ka}}$ and visual cluster $\{VC_{i}\}_{i=1}^{C_{kv}}$, where $C_{ka}$ and $C_{kv}$ represent the number of audio and visual clusters. Next, we compute the correlation between cluster centroids and generate approximate importance scores $P$ weighted by the number of tokens within each cluster, as shown below:
\vspace{-3pt}
\begin{equation}
\begin{aligned}
    \{AC_{i}\}_{i=1}^{C_{ka}}, I_{a} &= \mathrm{Kmeans}(\{ F_{a}^{i}\}_{i=1}^{N_a}, C_{ka}), \\
    \{VC_{i}\}_{i=1}^{C_{kv}}, I_{v} &= \mathrm{Kmeans}(\{\tilde{F}_{v}^{i}\}_{i=2}^{4}, C_{kv}), \\
    C_{ij} &= \frac{\mathrm{centroid}(AC_{i}) \times \mathrm{centroid}(VC_{j})^{T}}{\sqrt{D}}, \\
    P &= \frac{VC_{j}\exp(C_{ij})}{ {\textstyle \sum_{k=1}^{C_{kv}}}VC_{k}\exp(C_{ik}) },
\end{aligned}
\end{equation}
\vspace{-6pt}

\noindent where $I_{a}$ and $I_{v}$ represent the mapping from original tokens to their corresponding clusters. Considering that both audio and video contain redundant background information that interferes with foreground learning during segmentation, we employ the Top-p algorithm to filter importance scores and retain crucial audio-visual semantic associations. By combining the previously obtained cluster to token mappings $I_{a}$ and $I_{v}$, we project the semantic-level associations back to the spatial domain, yielding a correlation index matrix. Finally, we use this matrix with sparse matrix multiplication to compute attention between audio and video, achieving more accurate audio-visual fusion as follows:
\begin{equation}
\begin{aligned}
    Index &= \mathrm{SP}(\mathrm{TopP}(P,p), I_{a}, I_{v}), \\
    AM &= \mathrm{SMM}(\{ F_{a}^{i}\}_{i=1}^{N_a}, {\{\tilde{F}_{v}^{i}\}_{i=2}^{4}}^{T}, Index), \\
    out &= \mathrm{SMM}(\mathrm{Softmax}(AM), \{\tilde{F}_{v}^{i}\}_{i=2}^{4}, \mathrm{Index}),
\end{aligned}
\end{equation}

\noindent where $\mathrm{SP}(u, v, w)$ denotes the spatial mapping of $u$ based on the relationship between $v$ and $w$, $\mathrm{TopP}(u, p)$ represents filtering $u$ using the Top-P algorithm with probability $p$, and $\mathrm{SMM}(u, v, \mathrm{Index})$ indicates computing the sparse matrix multiplication between $u$ and $v$ only at valid positions specified by $\mathrm{Index}$.

\subsection{Asynchronous Dynamics Modulator}
\label{34}
To model the asynchronous and complex dynamic characteristics of audio and video, we propose a Mamba-based Asynchronous Dynamics Modulator (ADM). We examine Mamba's internal parameters and find that $\Delta$ directly controls the state transition speed in Mamba, determining how quickly information propagates through time. This property allows $\Delta$ to flexibly adapt to different temporal scales and capture the asynchronous nature of audio-visual dynamics. Our approach leverages Mamba's characteristics by first modeling video dynamics, then extracting audio dynamics and using them to control Mamba's $\Delta$ parameter, thereby achieving synchronized perception of audio-visual dynamics. The structure of ADM is illustrated in Figure \ref{fig:model2}(a), consisting of our proposed Audio-Dynamically Modulated Mamba (AMM), feed-forward network (FFN), layer normalization, and residual connections. Our AMM follows the common Mamba block design from prior workes, as shown in Figure \ref{fig:model2}(b). Our core improvement lies in the Audio-Dynamically Modulated SSM (AM-SSM), as shown in Figure \ref{fig:model2}(c), which enables modeling of asynchronous audio-visual dynamics.

After object queries $\{Q_{t}\}_{t=1}^{T}$ are input into the AM-SSM, where $Q_{t} \in \mathbb{R}^{B \times N_{q} \times D}$ and $N_{q}$ denotes the query number, they undergo dimension transformation through a linear layer, which then splits into content-aware $\mathbf{B}$, $\mathbf{C}$, and $\Delta$ parameters. The $\Delta$ at this stage captures the internal dynamics of the video. For audio features $\{ F_{a}^{i}\}_{i=1}^{N_a}$, we first apply a linear layer to transform their dimensions to match those of the video features. Note that the video has already undergone cross-scan for four-directional flattening. To maintain consistency, audio features also require flattening. We then employ grouped convolution and MLP to extract dynamic variations in the audio, represented as gating values. After sigmoid activation, these values are multiplied with the audio features and added to the original $\Delta$. This enables the updated $\bar{\Delta}$ to simultaneously perceive both video and audio dynamics. Subsequently, we use $\bar{\Delta}$ to perform the SSM modeling process, as follows:
\begin{equation}
\begin{aligned}
    \mathbf{B}, \mathbf{C}, \Delta &= \mathrm{Split}(\mathrm{Linear}(\{Q_{t}\}_{t=1}^{T})), \\
    \tilde{F}_{a} &= \mathrm{CS}(\mathrm{Linear}(\{ F_{a}^{i}\}_{i=1}^{N_a})), \\
    \bar{\Delta} &= \Delta + \tilde{F}_{a} \times \mathrm{Sigmoid}(\mathrm{MLP}(\mathrm{GC}(\tilde{F}_{a}, k))),
\end{aligned}
\end{equation}

\noindent where $\mathrm{Split}$ denotes splitting along the channel dimension, $\mathrm{CS}$ denotes cross-scan, and $\mathrm{GC}(u,v)$ denotes performing grouped convolution on $u$ with $v$ groups. Subsequently, we feed the $\bar{\Delta}$ along with other parameters into the SSM in Equation \ref{eq:2} for computation, yielding refined queries $\bar{Q}_{t}$.

\begin{table*}
\caption{Quantitative evaluation of different models from related tasks on the AVISeg test set.}
\vspace{-10pt}
\label{tab:sota1}
\resizebox{.9\textwidth}{!}{
\begin{tabular}{c|ccc|ccc|ccccc}
\shline
Task & Method   & Reference  & Audio & FSLA & HOTA & mAP  & FSLAn & FSLAs  & FSLAm &  AssA & DetA \\

\hline
\multirow{6}{*}{VIS} 
& Mask2Former-VIS \cite{mask2former2}  & CVPR' 22 & \ding{55} & 29.75 & 52.03 & 28.66 & 0.00 & 25.47 & 36.37  & 64.49 & 43.33 \\

& TeViT \cite{tevit}   & CVPR' 22 & \ding{55} & 32.28 & 53.67 & 31.52 & 0.00 & 28.07 & 39.18  & 65.27 & 45.10 \\

& SeqFormer \cite{seqformer}   & ECCV' 22 & \ding{55} & 30.32 & 54.32 & 32.79 & 25.03 & 21.76 & 36.46  & 67.25 & 45.23 \\

& VITA \cite{vita}   & NeurIPS' 22 & \ding{55} & 38.04 & 57.48 & 36.25 & 15.04 & 27.98 & 47.45  & 69.86 & 48.96 \\

& DAVIS \cite{dvis}   & ICCV' 23 & \ding{55} & 23.99 & 49.12 & 19.83 & 14.61 & 24.83 & 24.69  & 63.51 & 40.11 \\

& LBVQ  \cite{lbvq}  & TCSVT' 24 & \ding{55} & 34.73 & 56.97 & 36.58 & 27.71 & 29.52 & 38.96  & 68.34 & 48.83 \\

\hline
\multirow{3}{*}{AVSS} 
& AVSegFormer \cite{AVSegFormer} & AAAI' 24 & \ding{51} & 35.66 & 55.74 & 35.72 & 18.58 & 27.51 & 43.08  & 67.13 & 48.51  \\
& COMBO  \cite{COMBO}     & CVPR' 24 & \ding{51} & 39.49 & 57.39 & 37.84 & 21.91 & 27.18 & 49.63  & 68.87 & 50.12 \\
& SDAVS  \cite{pk}     & TMM' 26 & \ding{51} & 41.50 & 60.30 & 39.80 & 31.08 & 28.31 & 51.46  & 69.82 & 52.11 \\

\hline

& AVISM \cite{avis} & CVPR' 25 & \ding{51} & 42.78 & 61.73 & 40.57 & \textbf{32.22} & 29.83 & 52.40 & 71.15 & 54.97 \\

& ACVIS \cite{acvis} & ICASSP' 26 & \ding{51} & 42.87 & 62.09 & 42.14 & 21.16 & 30.62 & 52.34 & 70.64 & 55.30 \\

\rowcolor{gray!20}
\multirow{-3}{*}{AVIS} & H2S (Ours) & -- & \ding{51} & \textbf{45.96} & \textbf{63.32} & \textbf{43.21} & 30.83 & \textbf{34.97} & \textbf{54.93} & \textbf{71.81} & \textbf{57.37} \\
\shline
\end{tabular}}
\vspace{-10pt}
\end{table*}

\subsection{Loss Function}
Following previous works~\cite{avis}, we adopt a joint loss composed of frame-level, video-level, and similarity terms:
\begin{equation}
\mathcal{L} = \lambda_{\text{frame}} \mathcal{L}_{\text{frame}} + \lambda_{\text{video}} \mathcal{L}_{\text{video}} + \lambda_{\text{sim}} \mathcal{L}_{\text{sim}},
\end{equation}

\noindent where $\lambda_{\text{frame}}$, $\lambda_{\text{video}}$, and $\lambda_{\text{sim}}$ are set to 1, 1, and 0.5, respectively.  
For frame-level supervision, we compute matching costs between frame queries and ground-truth masks using the Mask2Former~\cite{mask2former} formulation, followed by Hungarian matching~\cite{kuhn1955hungarian} as in DETR~\cite{detr}.  
For video-level supervision, we match video queries with object trajectories following IFC~\cite{ifc} to enforce temporal consistency.  
Additionally, a similarity loss~\cite{vis, vita} aligns frame and video queries in the embedding space, encouraging features of the same identity to be close while separating those of different identities.

\section{Experiments}
\label{sec:exp}

\subsection{Datasets \& Evaluation Metrics}

\textbf{Datasets.}
We evaluate our proposed method on the AVISeg dataset \cite{avis}, a large-scale benchmark for audio-visual instance segmentation. AVISeg contains 926 videos with diverse and challenging scenarios, covering 26 sound categories across ``Music'', ``Speaking'', ``Machine'', and ``Animal''. Each video is annotated with pixel-level sounding object masks using a semi-automatic SAM-based tool~\cite{sam} and refined manually for quality assurance. The dataset includes long videos (average 61.4s) and is officially split into 616, 105, and 205 videos for training, validation, and testing.

\noindent \textbf{Evaluation Metrics.}
We evaluate model performance using mean Average Precision (mAP)~\cite{vis}, Higher Order Tracking Accuracy (HOTA)~\cite{hota}, and Frame-level Sound Localization Accuracy (FSLA) \cite{avis}.  
mAP measures detection quality across trajectories, while HOTA jointly evaluates detection and association consistency.  
FSLA assesses frame-wise localization accuracy by checking whether the predicted categories, object counts, and IoU exceed a given threshold $\alpha$, and includes variants (FSLAn, FSLAs, FSLAm) for silent, single-source, and multi-source scenarios.

\subsection{Implementation Details}
Following prior work, we validate the effectiveness of H2S using ResNet-50 \cite{resnet} and Swin-L \cite{swin} as visual backbones pretrained on ImageNet-1K \cite{imagenet} and MS-COCO \cite{coco}, and VGGish \cite{vggish} as the audio backbone pretrained on AudioSet. In ASD, we set $N_a$ to 8. In HCM, we set $C_{kv}$ and $C_{ka}$ to 8, and configure $p$ as 0.7, 0.8, and 0.9 from top to bottom stages. We train our model on the AVISeg for 48,000 iterations using the AdamW \cite{adamw} optimizer with a base learning rate of 0.0001. The learning rate is reduced by a factor of 10 at iteration 32,000. The batch size is set to 2, with each training sample containing 5 frames. For data augmentation, we apply multi-scale augmentation and random horizontal flipping. During training, the shorter side of input frames is randomly scaled between 360 and 480 pixels while maintaining the aspect ratio. During testing, the shorter side is fixed at 360 pixels. All experiments are conducted on two RTX 6000 GPUs.

\begin{table}
\begin{center}
\caption{Quantitative comparison with AVISM \cite{avis} under different backbones and pretraining datasets. Here, Ba. denotes the visual backbone, and Pr. denotes the pretraining dataset.}
\vspace{-8pt}
\label{tab:sota2}
\setlength{\tabcolsep}{3pt}
\renewcommand{\arraystretch}{0.97}
\resizebox{\linewidth}{!}{
\begin{tabular}{c|c|c|ccc|ccc}
\shline
Method & Ba.    & Pr.  & Param.  & GFLOPs & FPS  & FSLA    & HOTA  & mAP   \\
\hline
\multirow{2}{*}{AVISM \cite{avis}} &  & IN & \multirow{2}{*}{66.1M} & \multirow{2}{*}{3049} & \multirow{2}{*}{29} & 42.78 & 61.73 & 40.57 \\
 &  & COCO         &&      &       & 44.42 & 64.52 & 45.04 \\

\multirow{2}{*}{ACVIS \cite{acvis}} &  & IN & \multirow{2}{*}{138.2M} & \multirow{2}{*}{2916} & \multirow{2}{*}{25} & 42.87 & 62.09 & 42.14 \\
 &  & COCO         &&      &       & 46.48 & 65.12 & 46.68 \\

\rowcolor{gray!20}
&    & IN          &&          &        & \textbf{45.96} & \textbf{63.32} & \textbf{43.21} \\
\rowcolor{gray!20}
\multirow{-2}{*}{H2S (Ours)}  & \multirow{-6}{*}{R-50} & COCO               &   \multirow{-2}{*}{70.7M}    & \multirow{-2}{*}{3122}    & \multirow{-2}{*}{26}    & \textbf{47.58} & \textbf{65.70} & \textbf{48.54} \\
\shline
AVISM \cite{avis} &  &  & 238M & 10920 & 17 & 52.49&71.13&53.46 \\
ACVIS \cite{acvis} &  &  & -- & -- & -- & 54.17 & \textbf{72.96} & 54.16 \\
\rowcolor{gray!20}
H2S (Ours) & \multirow{-3}{*}{Swin-L} & \multirow{-3}{*}{COCO} & 242M & 10987 & 15 & \textbf{55.06} & 72.27 & \textbf{55.38} \\
\shline
\end{tabular}}
\vspace{-15pt}
\end{center}
\end{table}

\subsection{State-of-the-Art Comparisons}
Following AVISM \cite{avis}, we conduct a comprehensive comparison of our method with existing related approaches, including six Video Instance Segmentation (VIS) methods \cite{mask2former2,tevit,seqformer,vita,dvis,lbvq}, three Audio-Visual Semantic Segmentation (AVSS) methods \cite{AVSegFormer,COMBO,pk}, and two Audio-Visual Instance Segmentation (AVIS) method \cite{avis,acvis}, as shown in the Table \ref{tab:sota1}. For fair comparison, all methods employ an ImageNet pretrained ResNet50 as the visual backbone. Additionally, we compare our method with AVISM under different visual backbones and pretraining datasets, as presented in the Table \ref{tab:sota2}. Notably, when using a COCO pretrained ResNet50 as the visual backbone, our method outperforms previous approaches, improving the mAP by 7.8\%. Furthermore, this significant +3.5 mAP improvement over the AVISM baseline comes with minimal computational overhead, adding only +2.4\% GFLOPs and a marginal reduction of 3 FPS when measured on 60-frame videos at 480$\times$360 resolution. This confirms that H2S achieves a highly practical balance between detection effectiveness and inference efficiency.
\begin{figure*}
    \centering
    \includegraphics[width=.9\linewidth]{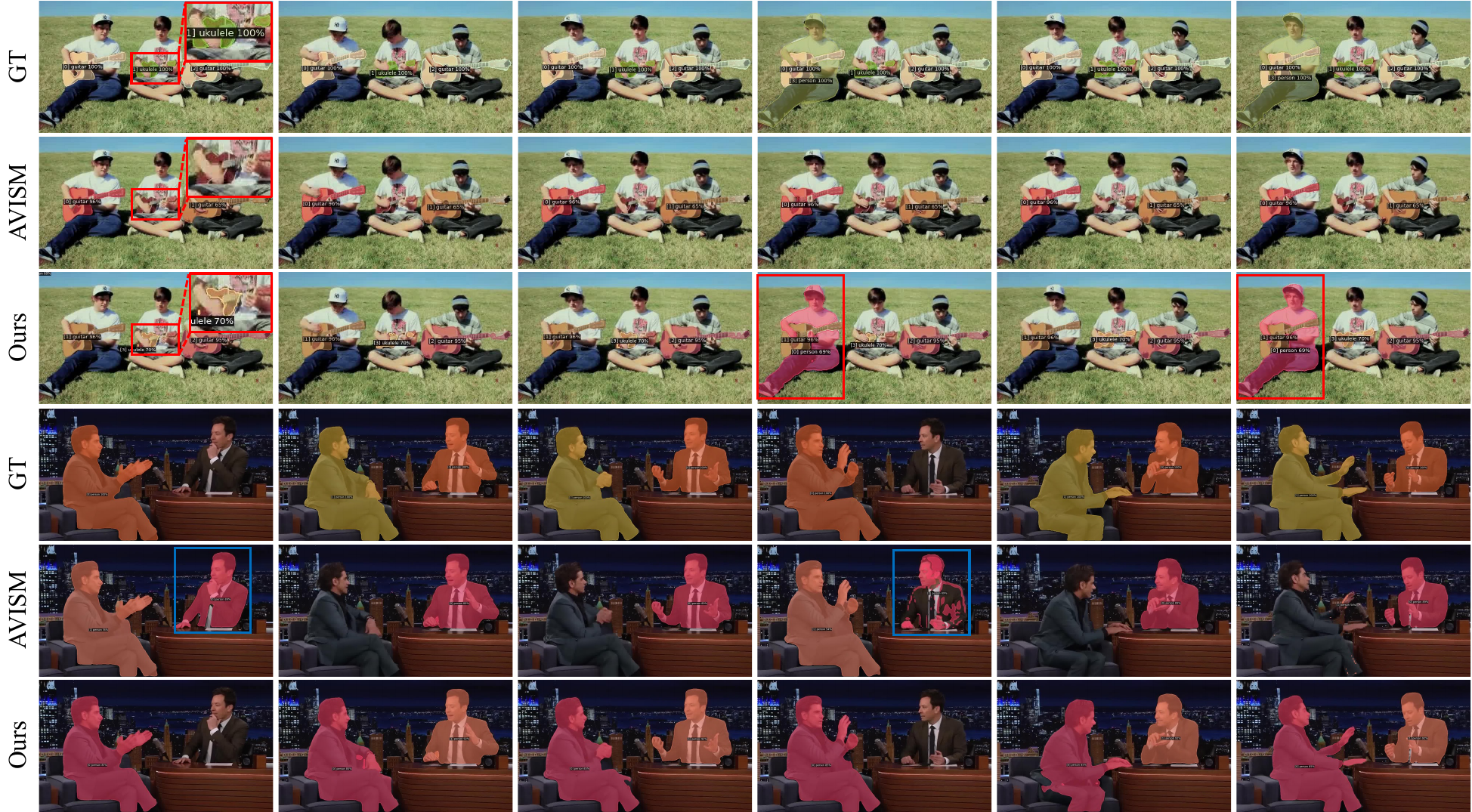}
    \vspace{-8pt}
    \caption{Visualization of segmentation results. Each group contains three rows: the first row shows the ground truth, the second row displays the predictions from AVISM \cite{avis}, and the third row presents the predictions from our method. Red boxes indicate sounding instances that should be segmented, while blue boxes denote silent instances that should be excluded.}
    \vspace{-10pt}
    \label{fig:fig}
    \Description{The figure illustrates the visualization of segmentation results.}
\end{figure*}

\begin{figure}
    \centering
    \includegraphics[width=.9\linewidth]{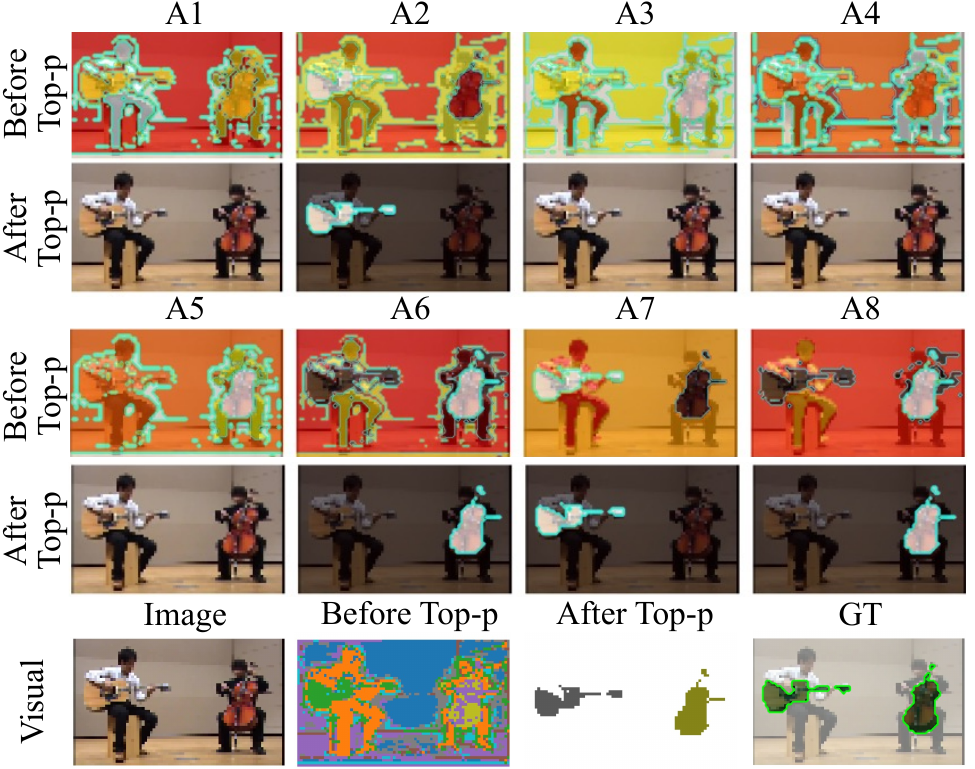}
    \vspace{-10pt}
    \caption{Visualization of the HCM. Correlation strength follows a specific color map. Rows 1-4 compare audio clusters before and after Top-p filtering. The bottom row demonstrates the spatial alignment achieved by visual cluster filtering alongside the ground truth.}
    \vspace{-12pt}
    \label{fig:topp}
    \Description{Visualization of the HCM. Correlation strength follows a specific color map. Rows 1-4 compare audio clusters before and after Top-p filtering. The bottom row demonstrates the spatial alignment achieved by visual cluster filtering alongside the ground truth.}
\end{figure}

\vspace{-5pt}
\subsection{How H2S Works: Disentangle Meets Dynamics}
To validate the effectiveness of our innovations, we visualize segmentation results as shown in Figure \ref{fig:fig}. We present two challenging scenarios involving multiple simultaneously sounding objects and state transitions. Each scenario contains three rows: ground truth, predictions from AVISM \cite{avis}, and predictions from our method. All methods use a COCO-pretrained ResNet50 as the visual backbone.
In the first scenario, when a ukulele and two guitars sound simultaneously, their audio signals are highly similar, with the ukulele's sound potentially masked by the guitars. Previous methods can only identify the dominant guitars in such acoustic overlap scenarios. However, our ASP first separates different sound sources through ASD, including the weaker ones, then establishes precise spatial correspondence via HCM. This enables accurate identification of the ukulele, as shown in the magnified view of the first column in the third row. When abrupt sound occurs, such as in columns 4 and 6 where a person suddenly vocalizes, previous methods fail to detect these state changes and miss the newly sounding instances. In contrast, our ADM modulates $\Delta$ to adapt to such dynamic variations: when sound states change abruptly, the model prioritizes current information to capture these transitions, successfully segmenting the suddenly sounding instances.
The second scenario is more challenging, presenting alternating vocalization with temporal asynchrony. When two people alternate vocalizing, both acoustic overlap and abrupt state changes occur. Previous methods may conflate overlapping audio as a single source, leading to missed instances. Additionally, when sound abruptly ceases, these methods remain insensitive to such state changes and continue segmenting silent instances, as indicated by the blue boxes. In contrast, our method successfully handles both challenges: ASP disentangles mixed audio and establishes precise cross-model correspondence to distinguish individual sources, while ADM adapts to dynamic changes, ensuring robust tracking throughout state transitions. These results validate the effectiveness of our approach.

\begin{figure}
    \centering
    \includegraphics[width=.9\linewidth]{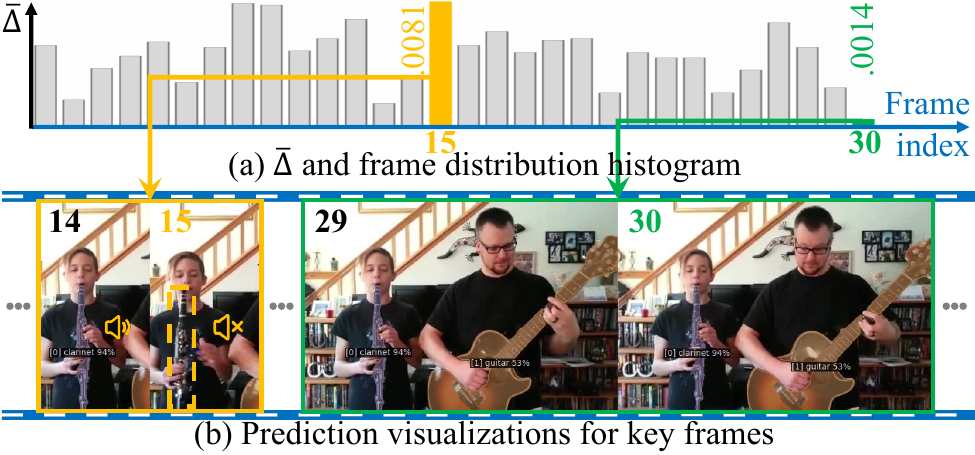}
    \vspace{-10pt}
    \caption{Visualization of the dynamic modulation $\overline{\Delta}$ and predictions under asynchronous audio-visual events. (a) Frame distribution histogram of the averaged $\overline{\Delta}$. (b) The model adaptively increases $\overline{\Delta}$ to capture abrupt state changes (frame 15) and decreases it during stable periods to maintain temporal continuity (frame 30).}
    \vspace{-12pt}
    \label{fig:rebuttal}
    \Description{The figure illustrates the structure of the Hierarchical Correspondence Mechanism (HCM).}
\end{figure}

\subsection{Refining Connections: The Power of Semantic Clustering}
We visualize the HCM to analyze the impact of Top-p filtering. Figure \ref{fig:topp} presents the semantic clustering and spatial alignment results. A specific color map represents the correlation strength. Dark regions indicate the background with zero correlation. Red areas represent medium correlation. Yellow or white regions highlight high intensity.
The first four rows display the correlation maps of eight distinct audio clusters. Each group compares the initial audio correlation with the refined results after Top-p filtering. Many initial clusters capture redundant background noise. Our Top-p filtering successfully suppresses these irrelevant distractions. A completely dark image after filtering indicates that the model discards the entire noise cluster. This refinement ensures the preservation of critical acoustic cues for cross-modal matching.
The bottom row illustrates the visual clustering and filtering process. The filtered visual clusters show remarkably highly accurate spatial alignment compared with the ground truth. These qualitative observations confirm that the ASP effectively filters irrelevant background noise. Our framework establishes precise and robust hierarchical correspondence between mixed audio and visual instances.

\subsection{Mastering Asynchrony: The Power of Dynamic Modulation}
\begin{figure}[t]
  \centering
  \begin{minipage}[t]{0.32\columnwidth}
    \vspace{0pt}
    \centering
    \includegraphics[width=\linewidth]{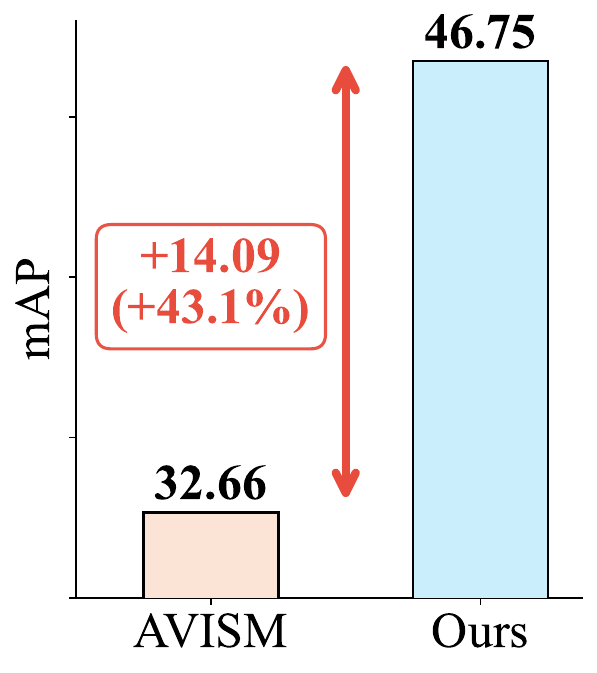}
    \vspace{-20pt}
    \caption{Performance comparison on the asynchronous subset.}
    \label{fig:im}
    \Description{This bar chart illustrates the performance on the asynchronous subset, highlighting that our H2S model significantly outperforms the AVISM baseline with a massive 14.09 mAP absolute increase, which is explicitly emphasized by a red vertical arrow and a text box indicating a 43.1 percent relative gain.}
  \end{minipage}\hfill
  \begin{minipage}[t]{0.65\columnwidth}
    \vspace{0pt}
    \centering
    \captionof{table}{Impact of different components in the H2S model.}
    \vspace{-10pt}
    \label{tab:ab1}
    \resizebox{\linewidth}{!}{
    \begin{tabular}{cccccc}
    \shline
    Index & ASP    & ADM   & FSLA    & HOTA  & mAP   \\
    \hline
    (1) & & & 43.11 & 63.49 & 43.90 \\
    (2) & \ding{51} & & 46.26 & 63.61 & 47.73 \\
    \rowcolor{gray!20}
    (3) & \ding{51} & \ding{51} & \textbf{47.58} & \textbf{65.7} & \textbf{48.54} \\
    \shline
    \end{tabular}}

    \vspace{8pt}
    \captionof{table}{Impact of the ASD. Here, CA denotes cross attention.}
    \vspace{-10pt}
    \label{tab:ab2}
    \resizebox{\linewidth}{!}{
    \begin{tabular}{ccccc}
    \shline
    Index & Settings   & FSLA    & HOTA  & mAP   \\
    \hline
    (1) & $w/o$ ASD & 45.69 & 65.34 & 47.01 \\
    (2) & $w/$ CA & 46.01 & 65.42 & 47.38 \\
    \rowcolor{gray!20}
    (3) & $w/$ ASD & \textbf{47.58} & \textbf{65.7} & \textbf{48.54} \\
    \shline
    \end{tabular}}
  \end{minipage}
\end{figure}

Real-world audio-visual events are often asynchronous. Sound can suddenly appear or disappear. To evaluate model robustness under these extreme conditions, we construct a challenging asynchronous subset. Our H2S framework achieves an impressive 46.75 mAP on this subset. This performance outperforms the AVISM baseline by a massive margin of 14.09 mAP, as shown in Figure~\ref{fig:im}. The baseline model relies on fixed temporal windows, treats sounding and silent states equally. Therefore, the baseline struggles with sudden state transitions. Our method overcomes this critical limitation through the proposed ADM. To explain this mechanism intuitively, we visualize the dynamic modulation parameter $\overline{\Delta}$ in Figure \ref{fig:rebuttal}. In our model, $\overline{\Delta}$ is generated for each object query. This parameter directly controls the state transition speed for individual instances. To observe the overall temporal dynamics, we calculate the average $\overline{\Delta}$ across all queries within each frame. Figure \ref{fig:rebuttal}(a) displays the frame distribution histogram of this averaged $\overline{\Delta}$.
Figure \ref{fig:rebuttal}(b) shows the corresponding video frames. As illustrated in the visualization, the sound abruptly ceases at frame 15. The model immediately detects this asynchronous change. The network adaptively increases the average value of $\overline{\Delta}$. This rapid increase allows the model to capture the sudden state transition. This adjustment prevents the model from tracking a silent object. Conversely, the acoustic environment becomes stable around frame 30. The model dynamically decreases the average value of $\overline{\Delta}$. This reduction helps the model prioritize historical continuity. Consequently, the system ensures robust instance tracking during stable periods. These observations provide strong evidence for our architectural design. Our module dynamically adjusts internal state transitions to maintain accurate tracking in complex scenarios.

\subsection{Ablation Study}
We conduct ablation studies on AVISeg using a COCO-pretrained ResNet as the visual backbone to evaluate the effectiveness of various components in our method.

\noindent \textbf{Impact of the ASP and ADM in the H2S model.}
We evaluate the impact of key components on H2S model performance, with particular focus on the Acoustic-Semantic Projector (ASP) and the Asynchronous Dynamics Modulator (ADM). As shown in Table \ref{tab:ab1}, adding either of these two modules consistently improves performance. For instance, incorporating the ASP increases FSLA by 3.15 and mAP by 3.83. Further introducing the ADM enhances HOTA by 2.09. These results validate our earlier claims: the ASP improves average precision and localization accuracy by establishing more precise correspondences, while the ADM enhances tracking accuracy by perceiving asynchronous dynamics. This comprehensive ablation study confirms the effectiveness of our model.

\noindent \textbf{Impact of the Audio Source Disentanglement.}
To verify the effectiveness of our introduced ASD, we design ablation experiments as shown in the Table \ref{tab:ab2}. Index (1) represents removing ASD, index (2) represents using cross attention to extract query information from audio features for interaction, and index (3) represents our method. The results demonstrate that although cross attention achieves certain performance gains, it still underperforms our explicit separation approach, validating our earlier claims. This ablation study confirms the effectiveness of ASD.

\section{Limitation \& Further Work}
\label{sec:lim}
Although our framework achieves SOTA performance in offline AVIS, it has not been validated in online settings, where models must perform inference without access to future data. Our design relies on full temporal context, which conflicts with online requirements. Extending our method to the online paradigm requires redesigning frame-level interaction mechanisms and video-level tracking procedures to ensure effectiveness under causal constraints. Specifically, a possible path for online adaptation involves replacing the audio separation module with online approaches that operate on streaming inputs using only past frames. To adapt the ADM to an online setting, we can incorporate a memory bank to preserve long-range historical features and employ causal scanning to avoid dependence on future information. We believe this represents an important direction for future research toward audio-visual understanding.

\section{Conclusion}
\label{sec:conc}
We present H2S, a framework for AVIS which addresses the key challenges of cross-model correspondence and asynchronous dynamics modeling. Our approach introduces two innovations: the ASP disentangles mixed audio into independent streams and establishes hierarchical correspondence with visual features, and the ADM leverages AMM to adaptively model temporal dynamics across modalities.
Experiments on the AVISeg dataset demonstrate that H2S achieves SOTA performance across multiple metrics. This work provides valuable insights for multimodel understanding tasks.


\begin{acks}
This work was supported in part by the National Natural Science Foundation of China under Grant 62372080, and Grant 62376050, and in part by the Natural Science Foundation of Liaoning Province under Grant 2024-MSBA-24. 
\end{acks}

\bibliographystyle{ACM-Reference-Format}
\balance
\bibliography{sample-base}

\clearpage
\appendix

\section{More Ablation Study}

\begin{table}
\begin{center}
\caption{Impact of hyperparameters in the HCM, including the number of clusters (left) and Top-P values (right).}
\vspace{-8pt}
\label{tab:ab3}
\resizebox{\linewidth}{!}{
\begin{tabular}{!{\hspace{-3pt}}c!{\hspace{-5pt}}c!{\hspace{-5pt}}c!{\hspace{-5pt}}c!{\hspace{-5pt}}c!{\hspace{-3pt}}c!{\hspace{-5pt}}c!{\hspace{-5pt}}c!{\hspace{-5pt}}c}
\shline
Index & $C_{ka}$, $C_{kv}$   & FSLA    & HOTA  & mAP & Top-P   & FSLA    & HOTA  & mAP   \\
\hline
(1) & 6 & 45.48 & 64.17 & 48.49& 0.7 & 45.38 & 64.27 & 47.47 \\
\cellcolor{gray!20}(2) & \cellcolor{gray!20}8 & \cellcolor{gray!20}\textbf{47.58} & \cellcolor{gray!20}\textbf{65.7} & \cellcolor{gray!20}48.54 & 0.8 & 47.03 & 65.28 & \textbf{48.67}\\
(3) & 10 & 44.88 & 64.07 & \textbf{48.87}& 0.9 & 45.56 & 64.35 & 47.79\\
(4) & [6, 8, 10] & 45.62 & 65.66 & 48.63& \cellcolor{gray!20}[.7, .8, .9] & \cellcolor{gray!20}\textbf{47.58} & \cellcolor{gray!20}\textbf{65.7} & \cellcolor{gray!20}48.54 \\
\shline
\end{tabular}}
\end{center}
\end{table}

\subsection{Impact of hyperparameters in the HCM.}
We conduct ablation experiments on the hyperparameters in HCM, as shown in Table \ref{tab:ab3}. For the number of clusters (left), indexes (1) to (3) set the cluster number to 6, 8, and 10, while index (4) sets them to 6, 8, and 10 from top to bottom stages. Setting the cluster number to 8 achieves the best performance, which may reflect the natural clustering of acoustic categories in the dataset. This validates the effectiveness of our semantic clustering strategy. For the top-p value (right), indexes (1) to (3) set the value to 0.7, 0.8, and 0.9, while index (4) applies 0.7, 0.8, and 0.9 from top to bottom stages. The results demonstrate that hierarchical configurations achieve optimal performance by applying stricter filtering to refined higher-level features and more inclusive learning to detailed lower-level features. These experiments confirm that our design choices in HCM effectively capture audio-visual correspondence at different semantic levels.

\begin{figure}
    \centering
    \includegraphics[width=.75\linewidth]{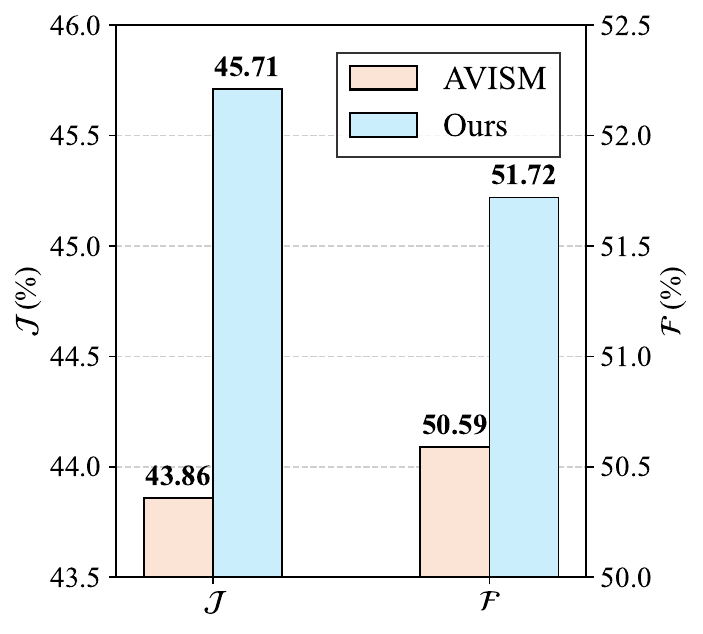}
    \vspace{-8pt}
    \caption{Zero-Shot Generalization on the AVSS dataset. The figure presents side-by-side bar charts comparing the AVISM baseline and our H2S framework. Metric $\mathcal{J}$ (left) and Metric $\mathcal{F}$ (right) utilize distinct Y-axis scales to clearly visualize the performance improvements. Our H2S model consistently outperforms the baseline across both class-agnostic metrics.}
    \label{fig:zero}
    \Description{This figure presents two side-by-side bar charts demonstrating that our H2S model consistently outperforms the AVISM baseline on the AVSS dataset across two class-agnostic binary metrics ($\mathcal{J}$ and $\mathcal{F}$), utilizing separated Y-axes to clearly highlight the performance gaps.}
\end{figure}

\subsection{Class-Agnostic Generalization on AVSS.}
To further verify the generalization performance of our H2S framework, we evaluate it on the large-scale AVSS dataset in a zero-shot setting. AVISM serves as the baseline for this cross-dataset comparison. Direct class mapping is challenging due to the significant disparity between AVIS (26 classes) and AVSS (70 classes). To ensure a fair evaluation, we convert the multi-class predictions into binary masks. We then evaluate the performance using class-agnostic binary metrics ($\mathcal{J}$ and $\mathcal{F}$). These binary metrics treat all sounding targets as a single foreground class. They allow us to evaluate the core audio-visual alignment capability without relying on specific class definitions.We present the quantitative results in Figure~\ref{fig:zero}. As illustrated in the figures, Metric $\mathcal{J}$ and Metric $\mathcal{F}$ utilize distinct Y-axis scales. This separation clearly visualizes the performance gap between the two methods. On Metric $\mathcal{J}$, H2S achieves 51.72, outperforming the AVISM baseline by 1.13. Similarly, on Metric $\mathcal{F}$, H2S reaches 45.71, marking a 1.85 improvement. These significant gains on an unseen dataset successfully demonstrate the robustness and adaptability of our proposed H2S model. They confirm that our model genuinely learns the fundamental correspondence between audio and visual signals.

\begin{table}
    \centering
    \caption{Impact of modulation variables in the ADM.}
    \vspace{-8pt}
    \label{tab:ab5}
    \resizebox{0.67\linewidth}{!}{
    \begin{tabular}{ccccc}
    \shline
    Index & Modulate   & FSLA    & HOTA  & mAP   \\
    \hline
    \rowcolor{gray!20}
    (1) & $\Delta$ & \textbf{47.58} & \textbf{65.7} & \textbf{48.54} \\
    (2) & $\mathbf{B}$ & 47.18 & 64.46 & 48.24 \\
    (3) & $\mathbf{C}$ & 46.97 & 64.17 & 47.98 \\
    \shline
    \end{tabular}}
    \vspace{-5pt}
\end{table}

\begin{table}
    \centering
    \caption{Effect of applying source separation.}
    \vspace{-8pt}
    \label{tab:ablation_source_separation}
    \resizebox{.8\linewidth}{!}{
    \begin{tabular}{lccc}
    \shline
    Methods & FSLA & HOTA & mAP \\
    \hline
    AVISM & 44.42 & 64.52 & 45.04 \\
    ~~+MixIT & 44.74 (+0.32) & 64.82 (+0.30) & 44.42 (-0.62) \\
    \rowcolor{gray!20}
    H2S (Ours) & \textbf{47.58 (+3.16)} & \textbf{65.70 (+1.18)} & \textbf{48.54 (+3.50)} \\
    \shline
    \end{tabular}}
    \vspace{-5pt}
\end{table}

\subsection{Impact of modulation parameters in the ADM.}
Unlike other methods that exchange $\mathbf{B}$ or $\mathbf{C}$ for multimodel fusion, we focus on the impact of $\Delta$ on AVIS. Since $\Delta$ directly controls the state transition speed in Mamba, it can flexibly adapt to different temporal scales of audio and video. We design a set of ablation experiments, as shown in Table \ref{tab:ab5}. Indexes (2) and (3) represent modulating $\mathbf{B}$ or $\mathbf{C}$, where the gated update term is added to $\mathbf{B}$ or $\mathbf{C}$ respectively. The experimental results show that modulating $\Delta$ achieves moderate improvement over modulating $\mathbf{B}$ or $\mathbf{C}$ on FSLA, a frame-level sound localization metric. In contrast, it demonstrates significant improvement on HOTA, a tracking accuracy metric. This validates our hypothesis and reflects that our design effectively captures temporal dynamic differences between audio and video.

\begin{table}
    \centering
    \caption{Ablation study on the MixIT pre-training datasets.}
    \vspace{-8pt}
    \label{tab:ablation_pretraining}
    \resizebox{.8\linewidth}{!}{
    \begin{tabular}{lccc}
    \shline
    Dataset & FSLA & HOTA & mAP \\
    \hline
    \rowcolor{gray!20}
    YFCC100M & 47.58 & 65.70 & 48.54 \\
    AVISeg & \textbf{47.81 (+0.23)} & \textbf{65.85 (+0.15)} & \textbf{48.62 (+0.08)} \\
    \shline
    \end{tabular}}
    \vspace{-5pt}
\end{table}

\subsection{Impact of Pre-training Datasets.} A potential concern is whether our performance gains merely stem from the massive external data used to pre-train the MixIT model. To address this concern, we evaluate the impact of different pre-training datasets for the audio separation module. As shown in Table \ref{tab:ablation_pretraining}, retraining the MixIT model exclusively on the target AVISeg dataset yields an mAP of 48.62. This result is highly comparable to the 48.54 mAP achieved using the default YFCC100M pre-trained weights. The performance differences between them are very marginal. Moreover, additional training on the target dataset requires a significant amount of extra time. Therefore, we choose to directly use the default pre-trained weights. These findings successfully rule out the influence of external data priors. They confirm that our performance improvements originate primarily from our proposed architectural designs.

\subsection{Effect of Audio Source Separation.} To verify the effectiveness of our source separation strategy, we compare our method with baseline approaches. As shown in Table \ref{tab:ablation_source_separation}, AVISM represents the baseline model. The +MixIT setting denotes the naive approach of directly concatenating separated audio features to the baseline. The results indicate that this naive integration actually degrades the baseline performance. It drops the mAP to 44.42. In contrast, our proposed framework successfully utilizes the separated acoustic signals to achieve 48.54 mAP. This demonstrates that explicit separation alone is insufficient. Our proposed ASP effectively establishes precise cross-modal correspondence. It successfully unlocks the potential of separated audio. These results clearly validate the effectiveness of our proposed method.

\section{More Limitation}
In this work, we aim to evaluate the impact of different audio source separation models. However, we face several practical challenges in achieving this goal. First, many existing separation models are not fully open-source. Some methods do not release their source codes. Other methods fail to provide their pre-trained weights or original training datasets. Second, the available open-source models lack a unified standard. Different methods evaluate their performance on completely different datasets. Their released weights are also optimized for different data distributions. These inconsistencies make a completely fair comparison extremely difficult. Finally, training all these separation models from scratch is a potential solution. However, this operation will introduce a massive amount of extra training time and computational cost. Therefore, we adopt the widely used MixIT model in our current framework. We identify the comprehensive evaluation of various separation models as an important direction for future work.

\end{document}